\documentclass[twocolumn,floatfix,altaffilletter,superscriptaddress,preprintnumbers,
               tightenlines,showpacs,showkeys,nofootinbib]{revtex4-1}
\usepackage[utf8]{inputenc}
\usepackage[colorlinks=true,citecolor=blue,linkcolor=blue]{hyperref}
\usepackage[normalem]{ulem}
\usepackage{amsmath,amssymb}
\usepackage{epsfig}
\usepackage{graphicx}               
\usepackage{url}
\usepackage{color}
\usepackage{slashed}
\usepackage{multirow}
\usepackage{placeins}
\usepackage[dvipsnames]{xcolor}
\usepackage{epstopdf}
\usepackage{soul}
\usepackage{tikz}
\usepackage[capitalise, english]{cleveref}
\usepackage{siunitx}
\usepackage{xspace}
\usetikzlibrary{trees}
\usetikzlibrary{decorations.pathmorphing}
\usetikzlibrary{decorations.markings}

\newcommand\myshade{80}
\colorlet{mylinkcolor}{ForestGreen}
\colorlet{mycitecolor}{Red}
\colorlet{myurlcolor}{violet}

\hypersetup{
  linkcolor  = mylinkcolor!\myshade!black,
  citecolor  = mycitecolor!\myshade!black,
  urlcolor   = myurlcolor!\myshade!black,
  colorlinks = true
}

\definecolor{jblue}{RGB}{20,50,100}
\definecolor{npurple}{RGB} {153, 51, 204}
\definecolor{wred}{RGB}{217,0,56}
\definecolor{white}{RGB}{255,255,255}

\definecolor{korange}{RGB}{235, 80,  43}
\definecolor{korange2}{RGB}{245, 100,  63}
\definecolor{kyelloworange}{RGB}{255, 210,  110}
\definecolor{kyelloworange2}{RGB}{240, 170,  90}
\definecolor{kred}{RGB}{204,  102, 153}
\definecolor{kpurple}{RGB}{153,  61, 190}
\definecolor{kpurplelight}{RGB}{213,  161, 230}

 \definecolor{tobycolour}{rgb}{.5,.0,.5}

\DeclareSIUnit\year{yr}
\DeclareSIUnit\pc{pc}
\DeclareSIUnit\ergs{ergs}
\DeclareSIUnit\msun{\ensuremath{M_\odot}}
\allowdisplaybreaks

\def\vev#1{\left\langle #1\right\rangle}

\def\thesection{\Roman{section}}

\makeatletter
\providecommand*{\diff}%
  {\@ifnextchar^{\DIfF}{\DIfF^{}}}
\def\DIfF^#1{%
  \mathop{\mathrm{\mathstrut d}}%
    \nolimits^{#1}\gobblespace}
\def\gobblespace{%
  \futurelet\diffarg\opspace}
\def\opspace{%
  \let\DiffSpace\!%
  \ifx\diffarg(%
    \let\DiffSpace\relax
  \else
    \ifx\diffarg[%
      \let\DiffSpace\relax
    \else
        \ifx\diffarg\{%
        \let\DiffSpace\relax
      \fi\fi\fi\DiffSpace}


\begin{document}

\title{Gravitational Imprints of Flavor Hierarchies}

\author{Admir Greljo}
\email{admir.greljo@cern.ch}
\author{Toby Opferkuch}
\email{toby.opferkuch@cern.ch}
\affiliation{Theoretical Physics Department, CERN, 1211 Geneva, Switzerland}

\author{Ben A. Stefanek}
\email{bstefan@uni-mainz.de}
\affiliation{PRISMA Cluster of Excellence and
             Mainz Institute for Theoretical Physics,
             Johannes Gutenberg-Universit\"{a}t Mainz, 55099 Mainz, Germany}

\date{\today}

\preprint{}

\begin{abstract}

The mass hierarchy among the three generations of quarks and charged leptons is one of the greatest mysteries in particle physics. In various flavor models, the origin of this phenomenon is attributed to a series of hierarchical spontaneous symmetry breakings, most of which are beyond the reach of particle colliders.  We point out that the observation of a multi-peaked stochastic gravitational wave signal from a series of cosmological phase transitions could well be a unique probe of the mechanism behind flavor hierarchies. To illustrate this point, we show how near future ground- and space-based gravitational wave observatories could detect up to three peaks in the recently proposed $PS^3$ model.

\end{abstract}

\maketitle

\section{Introduction}
The first direct detection of gravitational waves (GW)~\cite{Abbott:2016blz} was a stunning confirmation of the theory of general relativity and marked the discovery of the only messenger via which the universe can be probed back to the Planck era. To take advantage of this unique window into the universe, the next few decades will see a plethora of ground- and space-based gravitational wave observatories being built across twelve decades in frequency~\cite{Caprini:2015zlo,Yagi:2011wg,Kawamura:2011zz,Punturo:2010zz,Sathyaprakash:2012jk,Evans:2016mbw,Graham:2016plp,Graham:2017pmn}. In addition to what can be learned on the astrophysical front, this experimental effort offers an immense opportunity to probe fundamental physics in the early universe. Indeed, many particle physics processes that produce a stochastic gravitational wave background have already been identified, such as the primordial spectrum expected from inflation~\cite{Allen:1987bk,Turner:1990rc,Maggiore:1999vm}, violent first order phase transitions (FOPTs)~\cite{Weir:2017wfa,Curtin:2016urg,Katz:2014bha,Baker:2016xzo,Baldes:2018nel,Ellis:2018mja,Madge:2018gfl,Beniwal:2018hyi,Croon:2018kqn,Croon:2018erz,Brdar:2018num,Addazi:2018nzm,Breitbach:2018ddu,Angelescu:2018dkk,Alves:2018jsw,Kannike:2019wsn,Fairbairn:2019xog,Hasegawa:2019amx,Dunsky:2019upk,Athron:2019teq,DeCurtis:2019rxl,Addazi:2019dqt,Alanne:2019bsm}, cosmic strings~\cite{Vachaspati:1984gt,Damour:2004kw,Blanco-Pillado:2013qja,Dror:2019syi}, non-perturbative particle production~\cite{Khlebnikov:1997di,Easther:2006gt,Easther:2006vd,GarciaBellido:2007dg,GarciaBellido:2007af,Dufaux:2008dn,Machado:2018nqk}, primordial black holes~\cite{Sasaki:2018dmp,Barack:2018yly,Bartolo:2018rku}, etc. Many of these processes are expected to produce a GW spectrum with a single peak, with the notable exception being the nearly scale-invariant spectrum from inflation. 

Not as frequently discussed is the possibility of observing a multi-peaked gravitational wave signal, in either single or multiple experiments, and what such a signal might tell us about open puzzles in fundamental physics. One intriguing possibility is that a multi-peaked signal could come from a series of sequential FOPTs. As the peak frequency of the GW spectrum from a first order phase transition is set by the vacuum expectation value (VEV) in the broken phase, the observation of a multi-peaked signal could contain information about the scales of multiple spontaneous symmetry breakings (SSBs), with the first breaking giving the highest frequency peak and the last the lowest.

A longstanding question within fundamental physics is that of the flavor puzzle, which refers to why the Standard Model (SM) fermion Yukawa couplings are spread over so many orders of magnitude, with a top quark Yukawa that is $\mathcal{O}(1)$ but an electron Yukawa which is five orders of magnitude smaller. Just the quark sector alone has a hierarchy which covers 4-5 decades and also contains the puzzle of why the CKM mixing matrix is close to identity.
 
It has been proposed that the flavor hierarchies could be generated via a series of hierarchical SSBs~\cite{Grinstein:2010ve,Alonso:2013nca,Guadagnoli:2011id,Alonso:2011yg,Alonso:2013nca,Alonso:2012fy,Nardi:2011st,Espinosa:2012uu,Feldmann:2015zwa,Bishara:2015mha,Crivellin:2016ejn,Bordone:2017bld}. These types of models typically associate flavor with a fundamental gauge symmetry at high energies. The SM fermion masses and mixings are then generated via spontaneous breaking of this gauge symmetry, usually in several steps. The aforementioned models are compatible with the lowest SSB occurring at the TeV scale, which is highly motivated as it is the scale currently being probed at colliders (perhaps also in order to explain flavor anomalies~\cite{Lees:2013uzd,Hirose:2016wfn,Aaij:2015yra,Aaij:2014ora,Aaij:2017vbb,Aaij:2013qta,Aaij:2015oid,Aaij:2019wad,Buttazzo:2017ixm}). Interestingly enough, if this breaking occurs via a strongly FOPT, the resulting GW signal is in the sensitivity range of upcoming space-based interferometers such as LISA. Moreover, the higher breakings associated with light family mass generation may produce GW in the range of future ground-based interferometers such as Einstein Telescope (ET) and Cosmic Explorer (CE). Such a scenario would lead to a spectacular signature involving a multi-peaked GW signal, the peak frequencies of which contain information about the flavor hierarchies, spread across future GW experiments covering four decades of frequency space. This separation in frequency can be roughly seen by taking the geometric mean of the quark masses of each family, 
\begin{align}
 \nonumber
\sqrt{m_t m_b}\quad &:\qquad\sqrt{m_s m_c}\quad &&:\qquad\sqrt{m_u m_d} \\  \nonumber
1\quad &:\qquad 10^{-2} \quad &&:\qquad 10^{-4} \\  \nonumber
f^{-1}_\text{LISA}&:\qquad \dotso \quad &&:\qquad f^{-1}_\text{ET}	 \,.
\label{eq:sketchy_eq}
\end{align}
To further develop this idea, we will take the $PS^{3}$ model of Ref.~\cite{Bordone:2017bld} as a concrete example in what follows, though the concept generalizes to many models which solve the flavor puzzle through a series of hierarchical SSBs.

\section{Model Example: Pati-Salam Cubed}
\label{sec:PS3}

\begin{figure}
  \centering
\includegraphics[width=8cm]{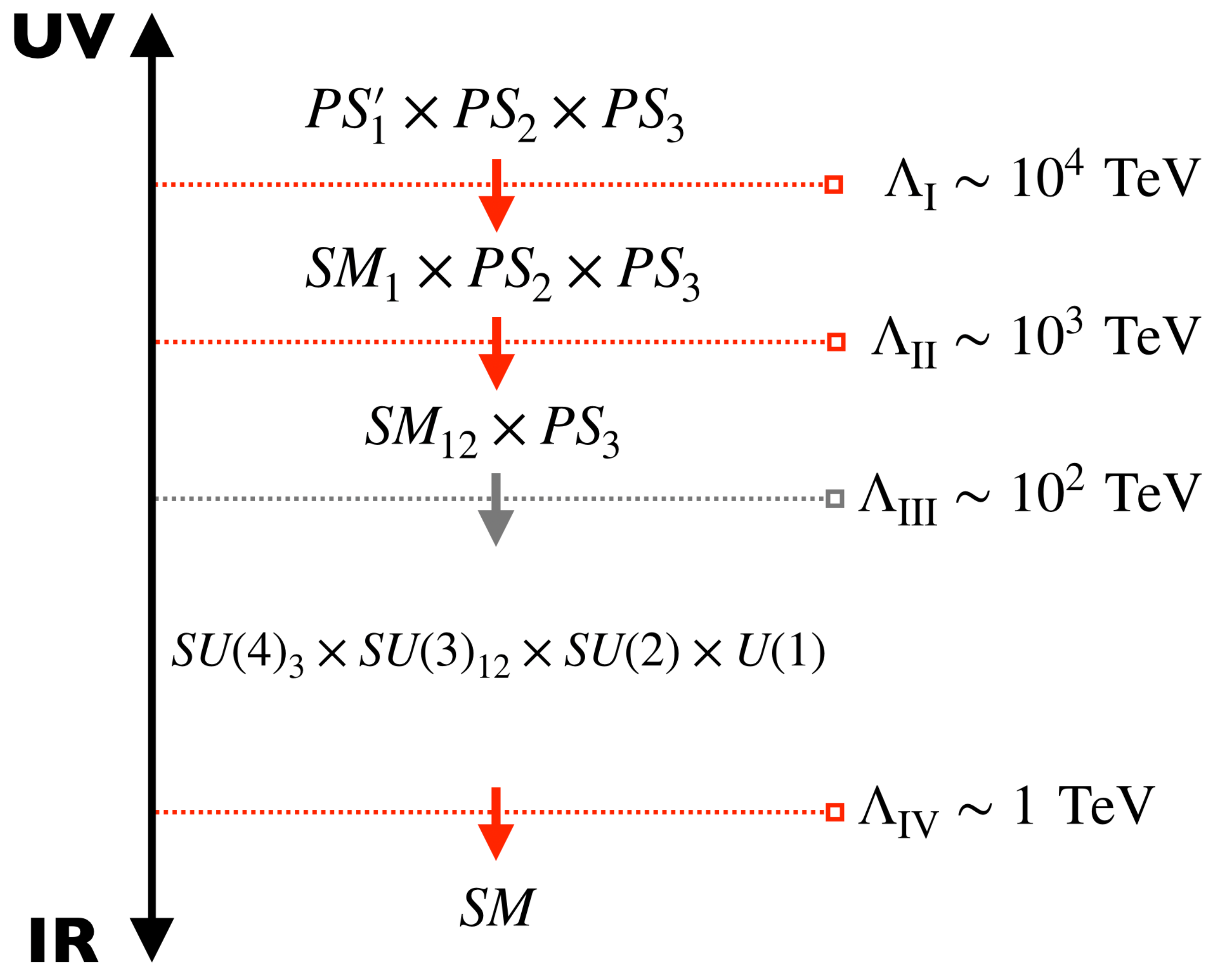}
  \caption{Schematic view of the $PS^{3}$ model detailed in \cref{sec:PS3}. Phase transitions marked with red arrows correspond to $SU(4)$ breakings (see \cref{sec:GWcalc}). }
  \label{fig:PS3_schematic}
\end{figure}

\begin{figure*}
  \centering
  \includegraphics[width=0.75\textwidth]{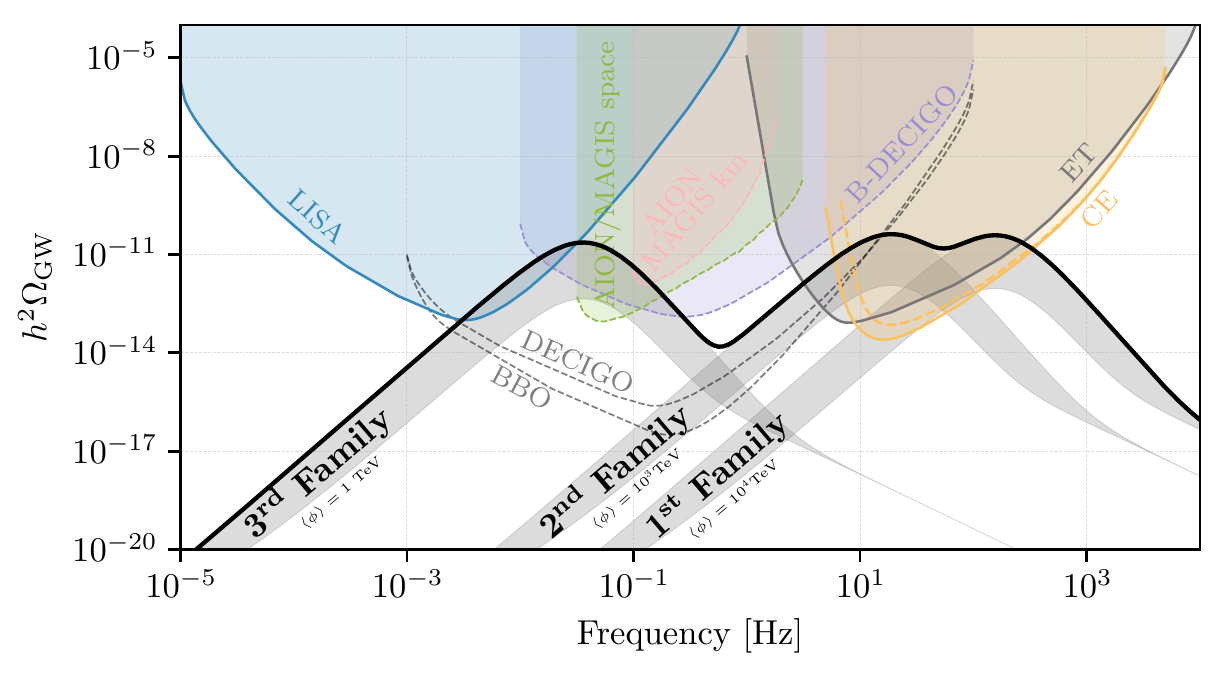}
  \caption{Complete GW spectrum, which we term the {\it Triglav signature}, following from three FOPTs in the $PS^{3}$ model. See \cref{sec:Results} for details.}
  \label{fig:Triglav}
\end{figure*}

As a prototypical example, we focus on the $PS^3$ model first introduced in Ref.~\cite{Bordone:2017bld}.\footnote{This is arguably the most compelling UV picture which offers a coherent explanation of the current flavor anomalies. See e.g.~\cite{DiLuzio:2017vat,Greljo:2018tuh,Bordone:2018nbg,Cornella:2019hct,Fuentes-Martin:2019bue}.} Here, the original Pati-Salam gauge group~\cite{Pati:1974yy} in higher-dimensional spacetime is deconstructed~\cite{ArkaniHamed:2001nc} onto three four-dimensional sites $PS^3 \equiv PS_1 \times PS_2 \times PS_3$, where each  copy acts on one family of SM fermions. In particular, an entire SM family including the right-handed neutrino fits into two left- and right-chiral multiplets, $\Psi^{(i)}_L \equiv ({\bf 4}, {\bf 2},{\bf 1})_i$ and $\Psi^{(i)}_R \equiv ({\bf 4}, {\bf 1}, {\bf 2})_i$, which embed quark and lepton doublets $Q^{(i)}_L$ and $L^{(i)}_L$ and singlets $u^{(i)}_R$, $\nu^{(i)}_R$, $d^{(i)}_R$ and $e^{(i)}_R$, respectively. The label $i = 1,2,3$ denotes the corresponding gauge group $PS_i \equiv [ SU(4) \times SU(2)_L \times SU(2)_R ]_i$.

The model undergoes a series of SSBs occurring at different energy scales as illustrated in Fig.~\ref{fig:PS3_schematic}. The first breaking after inflation is triggered by the VEV of $\Sigma_1$ in ${\bf 4}$ of $SU(4)_1$.\footnote{We propose a slight variation of the original model breaking $[SU(2)_R ]_1$ before inflation effectively solving the monopole problem of low-scale PS models~\cite{Jeannerot:2000sv}. $PS'_1$ in Fig.~\ref{fig:PS3_schematic} is defined as $[SU(4) \times SU(2)_L \times U(1)]_1$.} The subsequent breakings to the diagonal subgroups of neighboring sites are achieved by the appropriate scalar link fields in bifundamental representations, $\Phi^{L,R}_{i j}$ and $\Omega_{i j}$. More specifically,  $\Phi_{i j}$'s are in ${\bf 2}$ of the corresponding $SU(2)_i$ and $\bar {\bf 2}$ of $SU(2)_j$, while similarly, $\Omega_{i j}$ is $({\bf 4},{\bf 2},{\bf 1})_{i} \times (\bar {\bf 4},\bar {\bf 2},{\bf 1})_j$.   Finally, the Higgs fields live at the third site, e.g. $H_3 \equiv ({\bf 1},{\bf 2},\bar {\bf 2})_3$. 

Below the scale $\Lambda_{{\rm II}}$, the unbroken phase of the theory, $SM_{12} \times PS_3$, leads to an approximate $U(2)$ flavor symmetry observed in the SM at low-energies~\cite{Barbieri:2011ci}. The lower bound on this scale, $\Lambda_{{\rm II}} \gtrsim 10^3$~TeV, follows from stringent limits on flavor changing neutral currents (FCNC)  induced by the heavy gauge bosons coupling to the first two generations~\cite{Valencia:1994cj,Smirnov:2007hv,Kuznetsov:2012ai,Giudice:2014tma,Smirnov:2018ske}. At this level, Yukawa interactions are only allowed for the third family, $\mathcal{L} \supset \bar \Psi^{(3)}_L H_3 \Psi^{(3)}_R$, predicting vanishing light-fermion masses and a CKM matrix equal to identity. The smallness of neutrino masses is achieved by the inverse seesaw mechanism~\cite{Greljo:2018tuh}. The perturbation to this picture is obtained by higher-dimensional operators such as
\begin{equation}
\begin{split}
\mathcal{L}_{23} &= \frac{1}{\Lambda_{\rm III}}
 \bar \Psi^{(2)}_L \Omega_{2 3} H_3 \Psi^{(3)}_R + \textrm{h.c.}~,\\
\mathcal{L}_{12} &= \frac{1}{\Lambda_{{\rm II}}^2} \bar \Psi^{(k)}_L \Phi^L_{k 3} H_3 \Phi^R_{3 l} \Psi^{(l)}_R + \textrm{h.c.}~,
\end{split}
\label{eq:spurions}
\end{equation}
after the link fields acquire VEVs. The leading $U(2)$ breaking spurion, following from the first term, generates the mixing of the third and light families, $|V_{t s} | \sim \vev{\Omega_{2 3}} / \Lambda_{{\rm III}}$, where $\vev{\Omega_{2 3}} \sim \Lambda_{{\rm IV}}$.   The light fermion masses are instead due to the second term, with the largest being $y_{c} \sim \vev{\Phi^L_{2 3}} \vev{\Phi^R_{3 2}} / \Lambda_{{\rm II}}^2 $. Similarly, $y_{u}$ follows from $\Lambda_{\rm I}$, etc. The UV completion of the effective operators in Eq.~\eqref{eq:spurions} has been discussed in Refs.~\cite{Bordone:2017bld,Greljo:2018tuh}. We assume the scales generating the operators to coincide with the preceding symmetry breaking scales, e.g. $\Lambda_{{\rm III}} \sim \vev{\Phi_{2 3}}$ and $\Lambda_{\rm II}  \sim \vev{\Phi_{1 2}}$. From here it follows that the four-step breaking, $i)$ $10^4$~TeV, $ii)$ $10^3$~TeV, $iii)$ $10^2$~TeV, and $iv)$ $1$~TeV, is well compatible with the observed pattern of fermion masses and mixings at low-energies.\footnote{Another independent argument to keep the first two SSBs close to the bounds implied by FCNC is to avoid large tuning of the Higgs mass which is only partially screened from the first two sites.} As we will show later, the three $SU(4)$ phase transitions naturally induce a stochastic GW signature within the reach of next-generation interferometers.\footnote{The $SU(2)$ breakings at the scale $\Lambda_{{\rm III}}$ may also produce stochastic GW signatures, but lie in a suboptimal frequency range for LISA and ET. This provides additional motivation for proposed intermediate frequency experiments such as atom interferometers or DECIGO.}

While we work in the deconstructed four-dimensional picture, the higher-dimensional model relates the hierarchy of quark and charged lepton masses to the stabilization mechanism of branes in the bulk~\cite{Goldberger:1999uk}.
Additionally, the higher-dimensional gauge symmetry justifies small scalar quartic couplings~\cite{ArkaniHamed:2003wu} leading to an almost classically scale invariant potential which is crucial to ensure strongly FOPTs as shown later.


\section{Gravitational Wave Calculation}
\label{sec:GWcalc}

\subsection{Effective Potential}
\label{sec:toymodel}
To describe the first SSB in $PS^{3}$ at the scale $\Lambda_{{\rm I}}$, we calculate in a simplified $4 \to 3$ model where $SU(4)$ is broken to $SU(3)$ by the VEV of a complex scalar $\Sigma$ in the fundamental representation of $SU(4)$. The matter content includes one set of doublets $\Psi_{L}$ and $\Psi_{R}$, also in the fundamental representation of $SU(4)$. In the $PS^3$ model, scalar fields which break $SU(4)$'s have suppressed Yukawa interactions and scalar cross-quartics\footnote{The higher-dimensional operators generating Yukawa interactions in \cref{eq:spurions} give a negligible correction to the effective potential.}. As a result, the relevant part of Lagrangian for the GW calculation depends only on a few parameters. More explicitly,
\begin{equation}
\mathcal{L} = \overline{\Psi} i \slashed D \Psi - \frac{1}{4} (F^a_{\mu \nu})^2 + |D_\mu \Sigma|^2 + \lambda v^2 |\Sigma|^2 - \lambda |\Sigma|^4~\,,
\end{equation}
with $D_{\mu} = \partial_{\mu} -igA_{\mu}^{a}T^{a}$. Thus, the relevant parameters of the model are $g, \lambda,$ and $v$. The breaking $SU(4)\rightarrow SU(3)$ occurs when the complex scalar $\Sigma$ acquires a VEV of the form $\langle \Sigma \rangle = (0,0,0,v/\sqrt{2})^{T}$.  The 7 broken generators correspond to a massive vector leptoquark $U_{\mu}$ and $Z'$ gauge boson. The decomposition of $\Sigma$ under the unbroken $SU(3)$ is ${\bf 4} = {\bf 3+1}$, with the entire complex ${\bf 3}$ and the imaginary part of ${\bf 1}$ containing the leptoquark and $Z'$ goldstones, respectively. The remaining degree of freedom ${\bf Re} \, \Sigma_{4} \equiv \phi/\sqrt{2}$ is a massive radial mode. The full finite-temperature effective potential for $\phi$ is
\begin{equation}
V_{\rm eff}(g,\lambda,v,\phi,T) = V_{0} + V_{CW} + V_{T\neq 0}  \,,
\label{eq:effPot}
\end{equation}
where tree level potential $V_{0}$ is
\begin{equation}
V_{0}(\lambda, v, \phi) = -\frac{1}{2}\lambda v^{2} \phi^{2} + \frac{\lambda}{4} \phi^{4} \,.
\label{eq:treePot}
\end{equation}
The one-loop Coleman-Weinberg correction $V_{CW}$ is
\begin{equation}
V_{CW}(g,\lambda, v,\phi) = \sum_b n_b \frac{m_b^4(\phi)}{64 \pi^2} \left( \ln \frac{m_b^2(\phi)}{\mu_R^2} - C_a\right)~,
\end{equation}
which we have written here in Landau gauge using the $\overline {\rm MS}$ renormalization scheme which gives $C_{a} = 3/2$ (5/6) for scalars (gauge bosons). The sum on $b$ is over all bosons which have a $\phi$-dependent mass and $n_{b}$ is the total number of degrees of freedom of the boson. The final piece $V_{T\neq 0}$ is the finite temperature correction to the potential
\begin{equation}
V_{T\neq 0}(g, \lambda, v, \phi, T) = \frac{T^4}{2\pi^2}\sum_b n_b \,J_b\left(\frac{m^2_b(\phi) + \Pi_b(T)}{T^2}\right)\,,
\end{equation}
which includes a correction from resummed Daisy diagrams. The thermal function $J_{b}(x^{2})$, the $\phi$-dependent masses $m_{b}(\phi)$, and the Debye masses $\Pi_{b}(T)$ are all given in the supplemental material. As we will show later, in the $PS^3$ model with $g \sim \mathcal{O}(1)$ and small $\lambda$, $V_{T\neq 0}$ naturally induces a thermal barrier which leads to a strong FOPT.

The subsequent $SU(4)$ transitions at the scales $\Lambda_{\rm II}$ and  $\Lambda_{\rm IV}$ are modeled by the more complicated breaking pattern $SU(4) \times SU(3)' \rightarrow SU(3)$ which is presented in the supplemental material.

\subsection{Numerical Procedure}
The GW spectrum from a FOPT is described by four parameters \cite{Caprini:2015zlo,Turner:1992tz,Kamionkowski:1993fg,Grojean:2006bp}. These are the nucleation temperature $T_{\rm n}$ which describes the onset of the phase transition, the strength $\alpha$, the inverse timescale $\beta$, and the bubble wall velocity $v_{\rm w}$. Due to gauge bosons in the plasma with sizeable couplings to the bubble walls and a strongly FOPT, we are in the regime of non-runaway bubbles with  $v_{\rm w} \sim 1$ where the peak of the GW spectrum is determined by the sound wave contribution~\cite{Caprini:2015zlo}. The remaining parameters we compute from the effective potential in \cref{eq:effPot} using the \verb"CosmoTransitions"~\cite{Wainwright:2011kj} package, the results of which we have confirmed using our own code based on the method of Ref.~\cite{Espinosa:2018hue}. Thus, for a given set of model parameters $g,\lambda,v$ we compute the corresponding GW parameters $\alpha,\beta,T_{\rm n}$ which allows us to obtain the GW spectrum from a template function extracted from lattice simulation~\cite{Huber:2008hg,Hindmarsh:2015qta,Caprini:2009yp}. We then are able to perform a standard signal-to-noise ratio (SNR) analysis to determine the detectability of the signal, see e.g. Ref.~\cite{Thrane:2013oya}. More details can be found in the supplemental material.

\subsection{Results}
\label{sec:Results}

We show in \cref{fig:Triglav} a benchmark multi-peaked GW signal where the first two transitions would be detectable in ET/CE and the final TeV scale phase transition would be detectable in LISA. Remarkably, the predicted $PS^3$ symmetry breaking scales (\cref{fig:PS3_schematic}) correspond to peak frequencies in the optimal range for experiments. The solid black line is the total signal which is the combination of the individual spectra and corresponds to the nominal LISA recommendation for modeling GW formation and propagation~\cite{Caprini:2015zlo}. The gray bands correspond to a conservative treatment of the sound wave contribution which illustrates the amount of theoretical uncertainty.

In the benchmark signal of \cref{fig:Triglav}, different peaks are obtained by appropriately varying the VEVs ($ 1, 10^3, 10^4$~TeV), as well as the effective relativistic degrees of freedom in the plasma. The renormalization group evolution of $PS^3$ unambiguously determines the values of all gauge couplings at the relevant scales starting from the benchmark input $g_{4,3}(\Lambda_{{\rm IV}}) = 2$, $g_{4,2} (\Lambda_{{\rm II}}) = \sqrt{2}$, and matching to the strong coupling at the scale $\Lambda_{{\rm IV}}$.\footnote{The breaking of $SU(4)_i \times SU(3)_j \to SU(3)$ implies the matching condition $g_3^{-2} =  g_{4,i}^{-2} + g_{3,j}^{-2}$.} The $SU(4)$ coupling at the third site $g_{4,3}(\Lambda_{{\rm IV}})$ is chosen to be somewhat larger as suggested by the current flavor anomalies~\cite{Bordone:2018nbg}.  Finally,  the three quartic couplings are set to $\lambda (\Lambda_{{\rm IV}}) = 10^{-2}$,  $\lambda (\Lambda_{{\rm II}}) = 10^{-2}$, and $\lambda (\Lambda_{{\rm I}})= 0.5\times 10^{-2}$.

\begin{figure}
  \centering
  \includegraphics{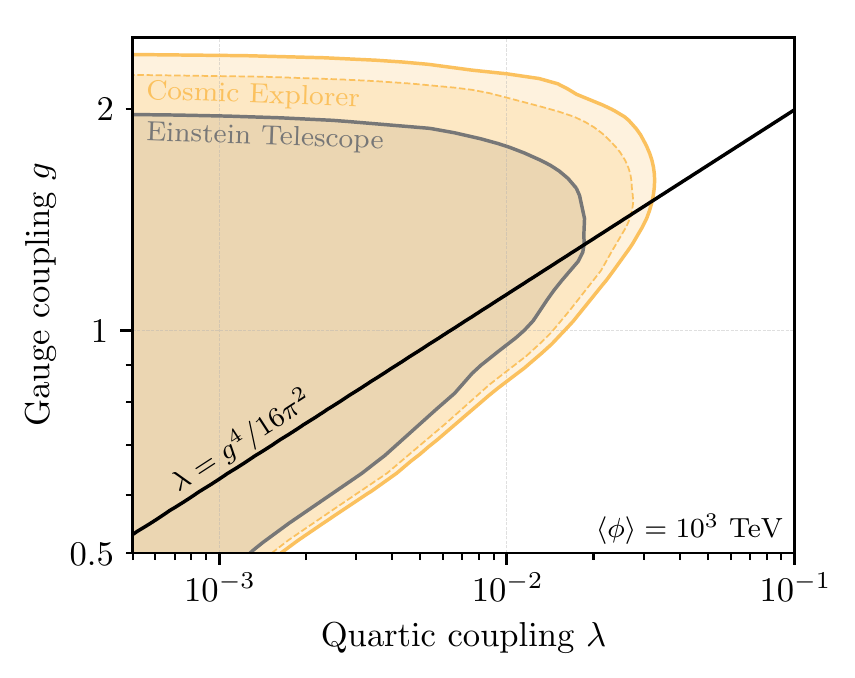}
  \caption{Detectability of GW from a FOPT in the $4 \to 3$ simplified model. See \cref{sec:Results} for details.}
  \label{fig:ToyModelDetectability}
\end{figure}

To assess how generic GW signatures are in $PS^{3}$, we show as an example in \cref{fig:ToyModelDetectability} the detectability of the GW spectrum computed in ET and CE, as a function of $g$ and $\lambda$ for a fixed VEV of $10^{3}$ TeV. These regions were computed using our simplified $4 \to 3$ model to calculate the GW parameters and spectrum, after which a detectability analysis is performed where we require an SNR of 5 to obtain the boundaries.  One can see immediately from \cref{fig:ToyModelDetectability} that significant parameter space exists which allows for a FOPT producing an observable GW signal without tuning. Furthermore, the best GW signatures are given for $i)$ gauge couplings of $\mathcal{O}(1)$ and $ii)$ small quartic coupling. 

Interestingly enough as discussed in \cref{sec:PS3}, both of these conditions are generic predictions of $PS^{3}$ because $i)$ it embeds the strong gauge group and $ii)$ the natural size of the quartic is set by the one-loop Coleman-Weinberg correction from the gauge sector. Indeed, the solid black line of \cref{fig:ToyModelDetectability} which falls nicely into the detectable region shows the expected size of $\lambda \approx g^{4} / 16\pi^{2}$ as would be generated from renormalization group flow. We have verified that $SU(4) \times SU(3)' \rightarrow SU(3)$ breaking pattern leads to qualitatively similar results.


\section{Conclusions}
\label{sec:discussion}

The peculiar pattern of hierarchical fermion masses which span many orders of magnitude is one of the longest standing puzzles in fundamental physics, the solution of which might require radical new approaches beyond colliders. In this letter we propose for the first time that a multi-peaked stochastic gravitational wave signature (where the ratios of peak frequencies follow the flavor hierarchies) could provide such a probe.

This idea is best illustrated in the context of the recently proposed $PS^{3}$ model for flavor hierarchies~\cite{Bordone:2017bld}, motivated also in part by the current $B$-meson anomalies. Here, the successful quark-lepton unification of the original Pati-Salam model is made compatible with flavor data by dimensional deconstruction onto three sites, one for each generation of SM fermions. 

We show that the parameters of the $PS^{3}$ model generically yield strongly first-order phase transitions as the gauge symmetry is sequentially broken down to the SM in hierarchical steps. Remarkably, the peak frequencies of the resulting GW spectra as determined by the VEVs fall precisely into the projected sensitivity range of future experiments. As we have argued, these are nearly inevitable predictions of the model as constructed. Such a spectacular signal, if observed, would offer a unique opportunity to probe the origins of the flavor hierarchies at energy scales which are currently inaccessible to colliders. 


\section*{Acknowledgments}

We thank Kfir Blum, Moritz Breitbach, Javier Fuentes-Mart\' in, Gino Isidori,  Andrey Katz, Joachim Kopp, Pedro Schwaller, and Marko Simonovi\' c for insightful discussions. TO has received funding from the European Research Council (ERC) under the European
Union's Horizon 2020 research and innovation programme (grant agreement No.\
637506, ``$\nu$Directions'') awarded to Joachim Kopp. 

\appendix 
\section{Model Details}
\label{sec:appendix_calc_detals}

\subsection{$SU(4) \to SU(3)$}

The last remaining pieces to complete the  $4\to3$ model of \cref{sec:toymodel} describing the $\Lambda_\text{I}$ transition is the determination of both the field dependent masses and Debye masses. The relevant fields entering the sum in \cref{eq:effPot} are $b = \phi, S_0, S, Z^{\prime\,T}_\mu, U_\mu^T, Z^{\prime\,L}_\mu, U^L_\mu$ with the corresponding number of degrees of freedom $n_b = 1, 1, 6, 2, 12, 1, 6$, while their field-dependent masses are: 
\begin{align}
	m_\phi^2 &= 3 \lambda \phi^2 - \lambda v^2\,, &\qquad
	m_{S, S_0}^2 &= \lambda \phi^2 - \lambda v^2\,,\\
	m_{Z'}^2 &= \frac{3 g^2 \phi^2}{8}\,,&\qquad
	m_{U}^2 &= \frac{g^2 \phi^2}{4}\,.
\end{align}
The Debye masses in the high-temperature limit are obtain following Refs.~\cite{Comelli:1996vm,Katz:2014bha,Laine:2016hma}, yielding
\begin{equation}
\begin{split}
\Pi_{\Sigma}(T) &= ~ \frac{5}{6} \lambda T^2 + \frac{15}{32} g^2 T^2\,,\\
\Pi^L_{A^a_\mu}(T) &= \frac{11}{6} g^2 T^2\,, \qquad \Pi^T_{A^a_\mu}(T)=0\,,
\end{split}
\end{equation} 
where $L$ and $T$ label the longitudinal and transverse components of the gauge field.

\subsection{$SU(4) \times SU(3) \to SU(3)$}

The SSBs at the scales $\Lambda_\text{II}$ and $\Lambda_\text{IV}$ are described by an $SU(4) \times SU(3)$ gauge group broken by the complex scalar field $\Omega_3 = (\overline {\bf 4},{\bf 3})$ to the diagonal $SU(3)$ subgroup. The decomposition of $\Omega_3$ under the unbroken $SU(3)$ is 
\begin{equation}
\begin{split}
\Omega _3^{i j} &= \frac{1}{\sqrt{3}} S_3 \delta_{i j} + \frac{1}{\sqrt{2}} O_3^a \lambda^a_{i j}~,\\
\Omega _3^{4 j} &= T_3^j~.
\end{split}
\end{equation}
where $i,j = 1,2,3$, and $\lambda^a_{i j}$ are the Gell-Mann matrices. $S_3$, $T_3^j$ and $O_3^a$ are a complex scalar singlet, triplet and octet, respectively. The real part of $S_3$ is the Higgs, while its imaginary part is the $Z'$ goldstone. The VEV of \textbf{Re}\,$S_3 =\phi / \sqrt{2} \equiv \sqrt{3} \, v_3/\sqrt{2}$ breaks the symmetry. $T_3$ is the goldstone of the vector leptoquark, while \textbf{Im} $O_3$ is the goldstone of the coloron. Finally, \textbf{Re} $O_3$ describes a physical scalar octet particle. In addition we require the following Dirac fermions: two copies of $(\mathbf{4},\mathbf{1})$ and two copies of $(\mathbf{1}, \mathbf{3})$ which embed the SM fermions. However, these fermions do not have Yukawa interactions with $\Omega_3$ and hence only contribute to the Debye mass for the gauge bosons. 

In what follows, we neglect contributions from the small quartic couplings everywhere except in the tree-level potential.
As a result we need only include the field-dependent and Debye masses for the vector bosons.\footnote{We also neglect scalar Debye masses as any field dependent contribution to the potential is always suppressed by the small quartic coupling.}

For the transverse polarizations of the gauge bosons the relevant field dependent masses are:
\begin{align}
m^2_{g'} &= \frac{(g_3^2 + g_4^2) \phi^2}{6}\,, \label{eq:coloron_mass}\\
m^2_{U_1} &= \frac{g_4^2 \phi^2}{12}\,, &\qquad m^2_{Z'} &= \frac{g_4^2 \phi^2}{24}\,, \label{eq:UZp_masses}
\end{align}
with corresponding degrees of freedom $n_b = 16,12,2$, respectively. For the longitudinal polarizations the Debye masses also play a role. These masses are:
\begin{equation}
\begin{split}
{SU(4):} \qquad \Pi^L_{H^\alpha_\mu}(T) &= \frac{13}{6} g_4^2 T^2\,,\\
{SU(3):} \qquad \Pi^L_{C^a_\mu}(T) &= 2 g_3^2 T^2 \,.
\end{split}
\end{equation} 
For the $U_1$ and $Z'$ vectors, one should simply sum the $SU(4)$ Debye mass and the field dependent masses in \cref{eq:UZp_masses} with $n_b^L=6,1$. For the coloron and gluon longitudinal modes we diagonalize the following matrix:
\begin{align}
m^2(\phi)+\Pi(T)&= 
\begin{pmatrix}
g_4^2\frac{\phi^2}{6} + \Pi^L_{H^\alpha_\mu}(T)  & -  g_3 g_4 \frac{\phi^2}{6} \\
- g_3 g_4\frac{ \phi^2}{6}  & g_3^2\frac{\phi^2}{6}   + \Pi^L_{C^a_\mu}(T)
\end{pmatrix}\,,
\end{align}
with $n_b^L = 8$. 

\section{Gauge coupling running}
\label{sec:gc_running}

Here we show that the renormalization group evolution of $PS^{3}$ unambiguously determines the values of all gauge couplings at the scales of interest once the values of the $SU(4)$ gauge coupling at the second and third sites $g_{4,2}(\Lambda_{\rm II})$ and $g_{4,3}(\Lambda_{\rm IV})$ are specified. This is because one must match onto the strong gauge coupling at the scale $\Lambda_{\rm IV}$
\begin{equation}
\frac{1}{g_{s}^{2}(\Lambda_{\rm IV})} = \frac{1}{g_{4,3}^{2}(\Lambda_{\rm IV})}  + \frac{1}{g_{3,3}^{2}(\Lambda_{\rm IV})} \,,
\end{equation}
which determines $g_{3,3}(\Lambda_{\rm IV})$ once $g_{4,3}(\Lambda_{\rm IV})$ is specified. This gives the initial condition for the running of the $SU(3)_{12}$ coupling, which is relevant in the phase of theory up to $\Lambda_{\rm II}$. At the scale of the second site $\Lambda_{\rm II}$, the coupling $g_{3,2}(\Lambda_{\rm II})$ is determined by an analogous matching condition after the specification of $g_{4,2}(\Lambda_{\rm II})$. The running of $SU(3)_{1}$ is then determined up to the scale of the first site $\Lambda_{\rm I}$, where the matching condition is simply $g_{3,1}(\Lambda_{\rm I}) = g_{4,1}(\Lambda_{\rm I})$. Thus, we have the values of all gauge couplings relevant for the GW calculation, namely the values of the $SU(4)$ and $SU(3)$ gauge couplings at the scales $\Lambda_{\rm I}$, $\Lambda_{\rm II}$, and $\Lambda_{\rm IV}$ where $SU(4)$'s are broken.

\begin{figure}
  \centering
  \includegraphics{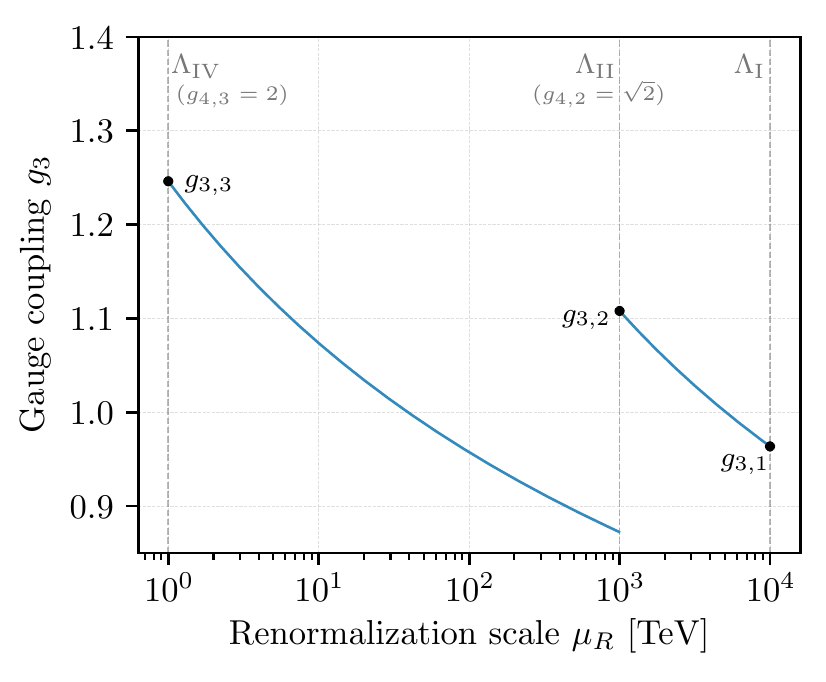}
  \caption{The running of the $SU(3)$ gauge coupling $g_{3}$ in different phases of $PS^{3}$. The input values $g_{4,2}(\Lambda_{\rm II}) = \sqrt{2}$ and $g_{4,3}(\Lambda_{\rm IV}) = 2$ correspond to the $B$-anomalies inspired benchmark point of \cref{fig:Triglav}.}
  \label{fig:g3_running}
\end{figure}

We show in \cref{fig:g3_running} an example running of the $SU(3)$ gauge coupling $g_{3}$ in the relevant phases of the theory, where we have taken $g_{4,3}(\Lambda_{\rm IV}) = 2$ as suggested for compatibility with flavor anomalies. The running of $g_{3}$ was determined using the one-loop beta function
\begin{equation}
(4\pi)^{2}\beta_{g_{3}} = -\mathcal{C} g_{3}^{3} \,,
\end{equation}
where $\mathcal{C}$ is a coefficient depending on the matter content. In the phase of the theory between $\Lambda_{\rm IV}$ and $\Lambda_{\rm II}$, the relevant matter content is that of Ref.~\cite{Greljo:2018tuh}. Specifically, we include two pairs of Dirac fermions in the fundamental representation of $SU(3)_{12}$ in addition to a complex scalar in $(\bar{{\bf 4}},{\bf 3})$, resulting in $\mathcal{C} = 23/3$. When running from $\Lambda_{\rm II}$ to $\Lambda_{\rm I}$ the matter content is the same except with only one pair of Dirac fermions charged under $SU(3)_{1}$, yielding $\mathcal{C} = 27/3$.

\section{Gravitational wave spectrum}
\label{sec:GW_spectra}

Before turning to the calculation of the GW spectrum we must first specify the last remaining ingredient to calculate $\alpha$, namely the effective number of relativistic degrees of freedom, $g_\star$, in the plasma. In $PS^3$ this is dominated by the new gauge degrees of freedom plus the usual SM fields. We find $g_*^{\text{I}} \simeq 250$, $g_*^{\text{II}} \simeq 220$ and $g_*^{\text{IV}} \simeq 150$ for the three transitions.

To determine the GW spectra from the particle physics outputs, namely $\alpha$, $\beta/H$ and $T_\text{n}$ as well as the choice of bubble wall velocity $v_\text{w}$, we utilize the recommendations from the LISA cosmology working group \cite{Caprini:2015zlo}. Here, fitting functions are given for the GW spectrum based on simulations of the plasma and bubble walls during the phase transition \cite{Huber:2008hg,Hindmarsh:2015qta,Caprini:2009yp}. Based on these recommendations we furthermore adopt the conventions of Ref.~\cite{Breitbach:2018ddu}, implementing the GW spectrum calculation using Tab.~(1) and Eqs.~(19) to (26) thereof. The case relevant in the scenario considered here is GW production in a thermal plasma, that is the sound wave and turbulence contributions dominate over the scalar field contribution.  

We include a suppression of the sound wave contribution given in Ref.~\cite{Breitbach:2018ddu}, as advocated in Refs.~\cite{Hindmarsh:2017gnf,Ellis:2018mja}. This is relevant for transitions that occur faster than a Hubble time, reducing the time acoustic sound waves have to source GWs. We use the following suppression factor $H \tau_\text{sw}$ in the expression for $\Omega_{\rm GW}$
\begin{align}
	H\tau_\text{sw} &= \text{min}\left\{1,\frac{4\pi^{1/3}}{\sqrt{3}} v_\text{w}\frac{H}{\beta} \left(\frac{\alpha\,\kappa_\text{sw}(\alpha)}{1+\alpha}\right)^{-\frac{1}{2}}\right\}\,, 
\end{align}
where the efficiency factor of the sound waves, $\kappa_\text{sw}(\alpha)$, is given in Eq.~(26) of Ref.~\cite{Breitbach:2018ddu}. The grey envelope of the GW spectrum shown in \cref{fig:Triglav} is produced with and without this suppression factor, corresponding to the lower and upper boundaries, respectively. As the sound waves are the dominant GW production mechanism, we expect this will correspond to the dominant source of uncertainty on GW production.	

The final ingredients are the experimental noise and power-law integrated curves used in the signal-to-noise ratio (SNR) analysis. Again we follow Ref.~\cite{Breitbach:2018ddu} closely, adding CE noise curves \cite{Evans:2016mbw}, for both pessimistic and optimistic projections, as well as the proposed atom interferometer experiments \cite{Graham:2016plp,Graham:2017pmn}, where we have taken the power-law integrated sensitivity curves from Ref.~\cite{Ellis:2019oqb}. To determine the experimentally accessible parameter space in  \cref{fig:ToyModelDetectability} we use an SNR threshold $\rho_\text{th}=5$ for both ET and CE.

\bibliographystyle{JHEP}
\bibliography{refs}

\end{document}